# Dynamical Classification of Supercooled Liquids: Critical Cooling Rates and Entropic Signatures

**B. Zhang[a], D. M. Zhang[b], D. Y. Sun[a§] and X. G. Gong[b§]**

[a]School of Physics and Electronic Science, East China Normal University, 200241 Shanghai, China

[b]Key Laboratory for Computational Physical Sciences (MOE), Institute of Computational Physics, Fudan University, Shanghai 200433, China

## Abstract

Using molecular dynamics simulations, we systematically investigate supercooled liquids formed at cooling rates below and above the critical cooling rate (CCR). By analyzing the distribution of short-time averaged potential energies (DoPE) and crystallization behaviors, we identify two distinct dynamical regimes in supercooled liquids: the glass-forming regime (GFR) and the crystal-forming regime (CFR). For systems cooled below CCR (CFR), the DoPE exhibits a sharp peak, indicative of reduced configurational entropy. In contrast, liquids cooled above CCR (GFR) display a broad DoPE distribution, reflecting higher configurational entropy. These findings establish a robust classification framework for supercooled liquids. Further analysis reveals a crossover temperature ($T_x$) in both regimes, consistent with the freezing temperature ($T_f$). Near $T_x$, crystallization barrier-temperature relationships exhibit abrupt changes. Below $T_x$, CFR crystallizes marginally faster than GFR, whereas above $T_x$, the influence of cooling rates on crystallization rates diminishes. These results further categorize GFR and CFR into high and low-temperature sub-regimes, highlighting the interplay between thermodynamics and kinetics in supercooled liquids.

[§]Corresponding Authors: xggong@fudan.edu.cn; dysun@phy.ecnu.edu.cn

# I. Introduction

In the field of glass research, the existence of a critical cooling rate (CCR) is an undisputed fact [1-9]. When the cooling rate falls below the CCR, the supercooled liquid experiences a first-order phase transition, leading to crystallization. Conversely, when the cooling rate surpasses the CCR, the supercooled liquid undergoes vitrification, resulting in the formation of glass [10-18]. The origin of the CCR has been attributed to the competition between two characteristic timescales: the available relaxation time $\tau_W$ and the intrinsic relaxation time $\tau_R$. These models (hereafter labeled as the two-time model) propose that as the cooling rate increases, $\tau_W$ decreases, diminishing the opportunity for the system to revert to equilibrium [19-21]. According to this perspective, vitrification seems to be a purely kinetic phenomenon. However, if this statement holds true, it requires that there is no fundamental distinction between the supercooled liquid formed below and above the CCR. To our best knowledge, this assumption has been the subject of limited research. Without a doubt, the validity of this assumption is pivotal: it determines whether vitrification should be regarded as a purely kinetic issue or as a complex interplay between thermodynamics and kinetics. In our previous work [22], we found that, the CCR classifies supercooled liquids into two distinct regimes: glass-forming regime (GFR) and crystal-forming regime (CFR). Specifically, supercooled liquids formed below the CCR correspond to the CFR, while those formed above the CCR correspond to the GFR. The fundamental difference between these two types of supercooled liquids lies in the nature of atomic motion: in the GFR, atomic motion may exhibit stronger correlations compared to that in the CFR. The aim of this paper is to further investigate the differences between the GFR and CFR, focusing on both their thermodynamic properties and crystallization behaviors.

From a thermodynamic perspective, the distinction between GFR and CFR should stem from the accessible regions of the potential energy landscape (PEL) [23-25]. Therefore, GFR and CFR can be scrutinized through the lens of their respective PELs. The information about accessible PEL should be manifested in in the distribution of average potential energies (DoPE). The DoPEs $(\rho(E))$ allow us to examine the differences between GFR and CFR in terms of their configurational entropy and average energy [26-28].

From a kinetic standpoint, for any supercooled liquid, the crystallization will inevitably occur over a sufficiently long period of time [29-34]. Thermodynamically, the fluctuations enable the supercooled liquid to surmount the nucleation barrier and initiate the nucleation [35-39]. Thus, the rate of nucleation provides direct insights into the nucleation barrier. Considering the PEL of the corresponding crystal is well define, data on nucleation rates also yields insight into the accessible PELs of supercooled liquids.

In this study, we conducted systematic molecular dynamics (MD) simulations of copper and aluminum supercooled liquids at various cooling rates. By analyzing thousands of samples, we discovered that the DoPE of GFR is significantly flatter than that of CFR, leading to a higher configurational entropy for GFR. Through the analysis of crystallization dynamics, we identified a characteristic temperature, denoted as $T_x$, at which the relationship between nucleation barrier and temperature exhibits a sudden change.

## II. Computational Detail

The embedded atom method (EAM) potential was used to model interactions between copper atoms [40]. The many-body potential was adopted to characterize the interactions between atoms in aluminum [41]. The LAMMPS code [42] with a time step of 2 femtoseconds was employed for MD simulations. The simulation consisted of 2048 atoms in a cubic box with periodic boundary conditions applied in three directions.

GFRs exist in the temperature range from the glass transition temperature ($T_g$) to the melting point ($T_m$). While, CFRs can exist in a shorter temperature range, i.e., between the freezing temperature ($T_f$) and $T_m$. $T_f$ is defined as the temperature, at which crystallization initiates during uniform cooling of the system. In contrast, $T_m$ is the specific temperature, at which the free energies of the crystal and liquid are equal. Owing to the potential barrier separating the solid and liquid, $T_f$ is generally observed to be lower than $T_m$. To help readers to better understand the current work, Figure 1 provides a schematic illustration of the various supercooled liquids under investigation.

Four typical cooling rates ($10^{10}$K/s, $10^{11}$K/s, $10^{12}$ K/s and $10^{13}$K/s) are investigated for copper, and two typical cooling rates ($10^{10}$K/s and $10^{12}$K/s) are investigated for aluminum. According to previous study, the CCR for copper and aluminum is approximately $10^{12}$K/s and $10^{11}$K/s, respectively [22, 43-45]. The supercooled liquids formed at the cooling rates higher than CCR are associated with GFR, while that formed at the cooling rates lower than CCR correspond to CFR.

The systems were initially equilibrated at much higher temperature to ensure a well-defined liquid state. Then, the systems were cooled to 300K using the isothermal-isobaric ensemble (NPT) at zero pressure. For each cooling rate, a minimum of 300 independent NPT simulations are performed starting from different initial states. The extensive number of independent samples ensures thorough exploration of the accessible regions on PELs within the available simulation time. After the cooling process, at each temperature of interest, the long-time canonical ensemble (NVT) simulation was then engaged, in which the state obtained from NPT simulations is taken as the initial state. Therefore, there are also 300 independent NVT simulations at each temperature. Each simulation lasts for 50ns. With this protocol, the NVT simulations incorporate the effect of the cooling rate by inheriting the state information from the NPT ensemble. All data used to calculate DoPE and crystallization behaviors were taken from the NVT simulations.

In order to filter out the fluctuations that obscure the information from PELs, the short-time average of potential energy ($\bar{E}$) were calculated over every 10ns. The previous studies have shown that the short-time averaging technique is indeed effective in eliminating the effects of thermal fluctuations [46]. In the rest of this paper, the DoPE refers the distribution of $\bar{E}$. For samples crystallized during NVT simulations, only the time interval before crystallization occurs, as indicated by $\Delta t$ in the right panel of Figure 1, is used to calculate the DoPE. For those that do not crystallize, the DoPE is calculated over the entire NVT simulations.

Previously, many studies have probed the PEL of supercooled liquids through inherent structures (ISs) by rapidly annealing the system from relatively high temperature [24, 47-56]. Our approach diverges from this method. Firstly, we focus on the short-time average of potential energy at a given temperature, rather than the energy of ISs. The short-time average of potential energy provides an instant reflection of the PEL and includes the effect of entropies. Secondly, we include the ensemble average over many independent samples, reducing the error on final results induced by long-time relaxations. Thirdly, few previous studies have emphasized the issue on cooling rates. On the contrary, the current studies specifically examine the issue of cooling rates.

To calculate the crystallization ratio ($P_c$), we define a "crystallization observation windows ($w_t$)", as illustrated in the right panel of Figure 1. $P_c$ is defined as the proportion of crystallized samples within $w_t$, i.e., $P_c = \frac{n_c}{n_c+n_g}$ with $n_c$ and $n_g$ as the number of crystallizing and non-crystallized samples, respectively. Here $w_t$ is taken as 5ns. The crystallized samples are defined as those in which crystallization occurs within the crystallization observation windows. Conversely, samples in which crystallization does not occur within these windows are defined as non-crystallized samples. The schematic plot in the right panel of Figure 1 illustrates this classification. Specifically, if the energy does not exhibit a noticeable jump, the sample is considered as non-crystallized samples (shown as the blue line in the right panel of Figure 1). If the energy shows an evident jump, characteristic of a first-order liquid-solid phase transition, the sample is labeled as a crystallizing sample (shown as the red line in the right panel of Figure 1). The final value of $P_c$ is obtained by averaging the results across numerous crystallization observation windows, each characterized by the same $w_t$ but with distinct starting points. It should be noted that, the crystallized samples, referring to supercooled liquid that crystallizes during the extended NVT simulation, can be either CFR or GFR.

Similarly, we also calculate the survival probability ($Q(t)$) of the supercooled liquid as a function of time, which is defined as the proportion of non-crystallized samples at time $t$, namely $Q(t) = \frac{n_g(t)}{n_g(0)}$. Here $n_g(t)$ and $n_g(0)$ are the number of non-crystallized samples at time $t$ and 0. Using $Q(t)$, we can investigate how the crystallization barrier changes with temperature.

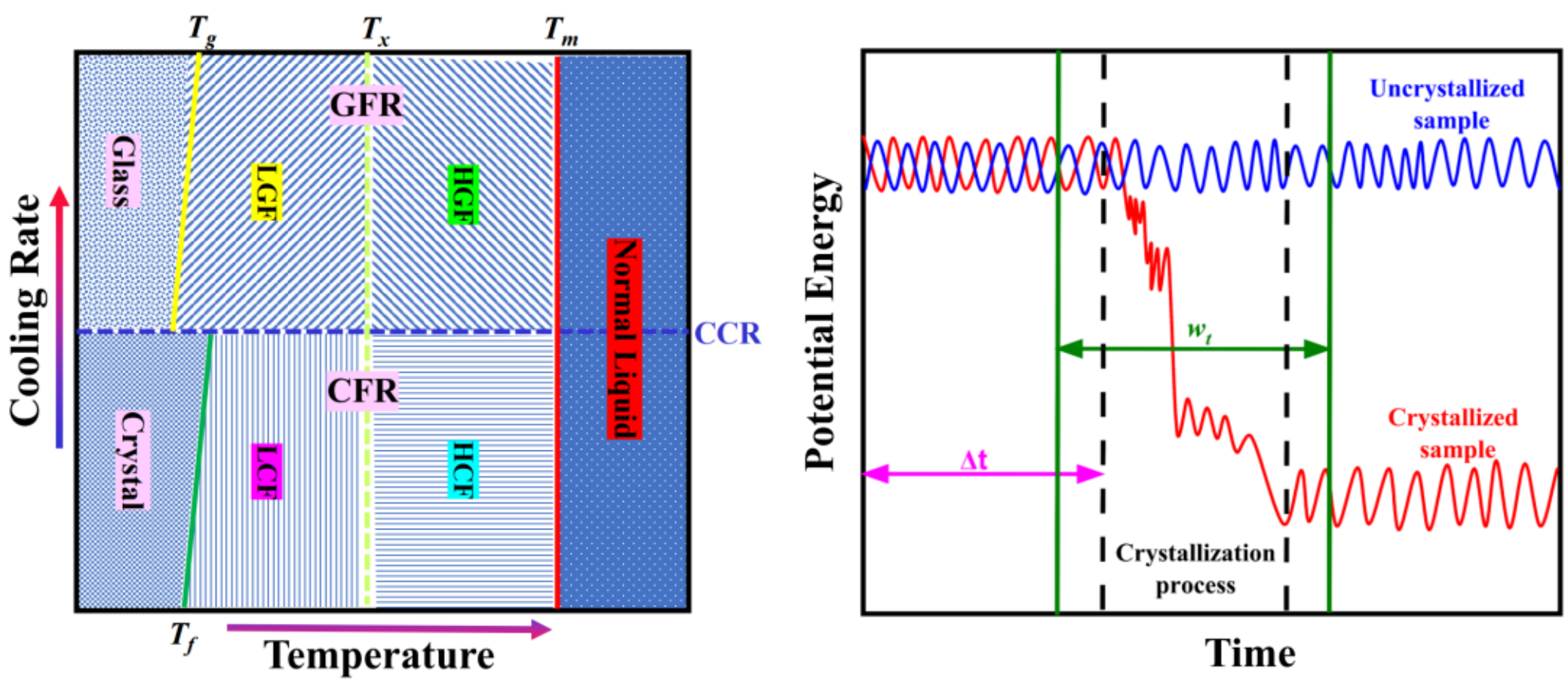

*Figure 1. Left panel: a schematic plot of the supercooled liquids under investigation. $T_m$, $T_g$ and $T_f$ refer the melting point, the glass transition temperature and the freezing temperature, respectively. The blue dash line denotes the critical cooling rate (CCR). The supercooled liquids formed above CCR were designated as GFR, while those formed below CCR were designated as CFR. $T_x$ is a characteristic time within both GFR and CFR identified in this work. $T_x$ further delineates both GFR and CFR into two distinct regions. For GFR, the high-temperature and low-temperature supercooled liquids are referred to as HGR and LGR, respectively. Similarly, for CFR, they are referred to as HCR and LCR, respectively. Right panel: schematic plot of the energy variation of supercooled liquids over time at a certain temperature. The blue line represents a uncrystallized sample, as indicated by the nearly constant average energy. The red line represents a crystallized sample, where the average energy begins to jump at certain time. $\Delta t$ is the time used to calculate the average potential energy for the crystallized sample. For uncrystallized samples, the average potential energy is calculated over the entire time.*

## III. Result and Discussion

### 1) Crystallization and Glass Transition

Figure 2 shows the potential energy as a function of temperatures for copper and aluminum. For GFR, the average potential energy exhibits a continuous change with temperature, reflecting a typical glass transition. The glass transition temperature $(T_g)$ can be determined by linearly extrapolating the variation of energy or volume with temperature at high and low temperatures. $T_g$ corresponds to the intersection of these two extrapolated lines. We find that, $T_g$ obtained based on energy is typically lower than that based on volume. For copper, $T_g$ is around $670K$ when determined from energy and $894K$ when determined from volume. For aluminum, $T_g$ is around $554K$ and $612K$ for energy and volume, respectively. In this paper, we adopt the average value of these two measurements as the representative $T_g$. In contrast, for CFR, $T_f$ is around $900K$ and $610K$ for copper and aluminum, respectively, as indicated by the energy or volume starting a sharp decrease in the average potential energy. It is noteworthy that, the value of $T_f$ and $T_g$ are averaged over more than 300 independent cooling samples. For an individual cooling process, both $T_f$ and $T_g$ can vary by as much as 30K.

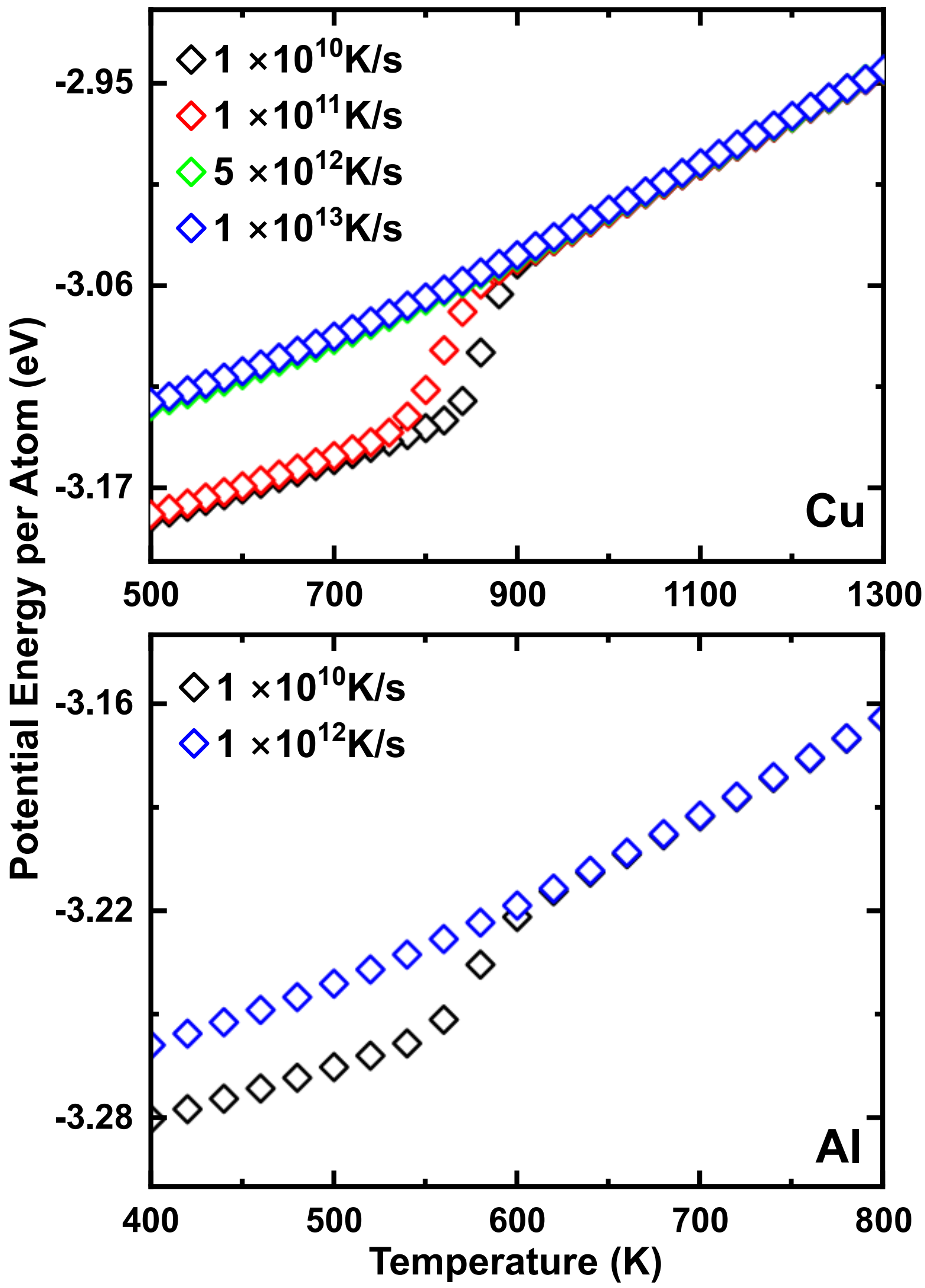

*Figure 2. Temperature dependence of the potential energy at different cooling rates for copper (upper panel) and aluminum (lower panel). For cooling rates higher than CCR, the system undergoes a glass transition. For cooling rates lower than CCR, the system crystallizes.*

As illustrated in Figure 2, despite significant variations in cooling rates, no substantial differences in potential energies are observed between GFR and CFR prior to $T_f$. In fact, as discussed in previous studies [22], conventional thermodynamic quantities and structural analysis are incapable of distinguishing between CFR and GFR. Similar to previous work, we also found that the cooling rate also exerts a negligible influence on the radial distribution function of supercooled liquids.

It is important to note that the data depicted in Figure 2 encompass not only time averages but also ensemble averages over at least 300 independent cooling samples. A number of previous studies have concentrated on the impact of cooling rates on the physical properties of supercooled liquids [14, 57-61]. These investigations have revealed that, in certain systems, cooling rates can significantly affect specific thermodynamic quantities of supercooled liquids. Our results indicates that, the impact of cooling rate on macroscopic thermodynamic quantities becomes markedly diminished when the ensemble averaging is taken into account.

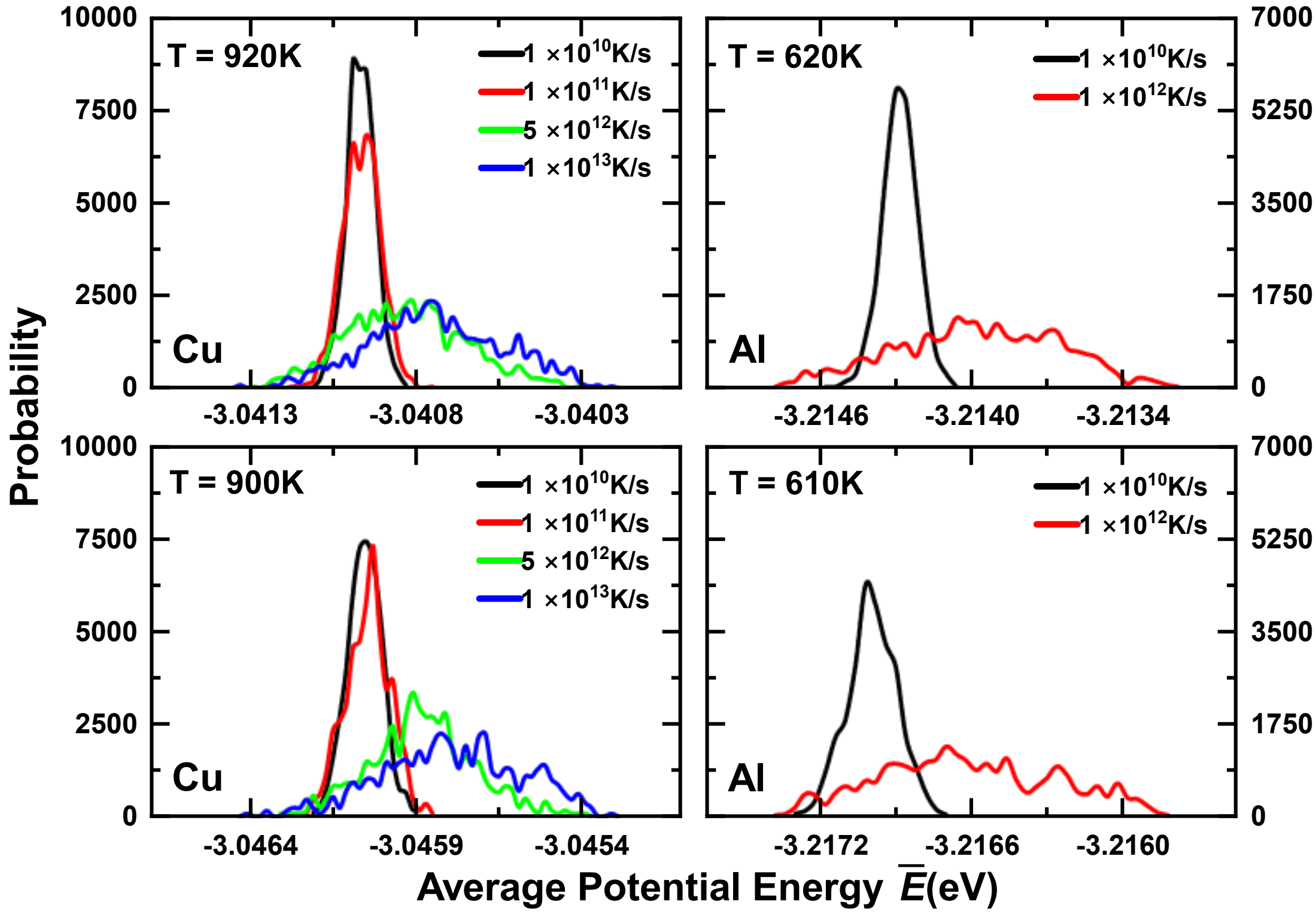

*Figure 3. Distribution of the average potential energy* $\bar{E}$ *of supercooled liquids at different cooling rates for copper and aluminum. The distribution of average potential energies of GFR is generally much broader than that of CFR, indicating that GFR may have a larger configurational entropy than CFR.*

**2) Distribution of Average Potential Energies**

Figure 3 illustrates the DoPE of CFR and GFR for both copper and aluminum at two different temperatures. From this figure, it becomes apparent that the DoPE for GFR is considerably broader, whereas CFR exhibits a pronounced main peak with only a minimal number of secondary peaks. By calculating the standard deviations of these two distributions, we find that the effective width of GFR is approximately four times that of the CFR. Given the fact that the DoPE is intrinsically linked to the topological structure of the accessible PEL, DoPE should fundamentally reflect the distribution of local minima on the accessible PEL. This observation suggests that CFR resides in a steep region of the PEL, while GFR is found in a relatively flat area. Accordingly, GFR should possess a larger configurational entropy compared to CFR. As depicted in Figure 3, it can be observed that the cooling rate exerts a slightly more pronounced impact on GFR than on CFR. This is mainly reflected in the shift of the main peak, rather than in the change of the distribution width. The much dissimilar behavior between GFR and CFR implies that the glass transition is not simply determined by kinetic factors, rather thermodynamic effect is also indispensable.

The vitrification is usually discussed based on the two-time model [19-21] mentioned above. We have noticed that, in using this model, the dynamics issue is often highlighted, while the thermodynamics is usually neglected. Our results indicate that, the competition between two characteristic times, the available relaxation time $\tau_W$ and the intrinsic relaxation time $\tau_R$, may exert a more significant influence on thermodynamic behavior. The significant differences in thermodynamics, in turn, may also affect the relaxation behavior, thereby further amplifying the effect of the cooling rate. Thus, we guess that, this two-time competition mechanism is responsible for the localization of CFR and GFR in two different regions of the PEL, thereby differentiating vitrification from crystallization.

As illustrated in Figure 3, the DoPE of both CFR and GFR exhibit a slight broadening as the temperature decreases. This phenomenon may be attributed to the fact that at higher temperatures, the system has a shorter relaxation time ($\tau_R$), which allows it to more readily explore and locate more stable regions on the PEL. Conversely, at lower temperatures, the system appears to become "trapped" on the PEL due to the scarcity of available stable regions. This is in stark contrast to the equilibrium distribution, where a higher temperature typically results in broader distributions.

Many previous studies have calculated the energy distribution of ISs in supercooled liquids [62-66]. The energy distribution of ISs closely resembles the standard Gaussian form. Although this distribution can change with the environment, the general Gaussian form remains largely invariant. The energy distribution of ISs observed in those studies markedly contrasts with the DoPE reported in the present work. This divergence suggests that the underlying physics governing these two distributions are distinct, and consequently, the insights they offer regarding the PEL are also different.

The Kauzmann paradox stands as a seminal issue in the study of supercooled liquids [28]. Based on the findings of this research, we anticipate that GFR would not persist across all cooling rates. Specifically, once the cooling rate dips below the CCR, the supercooled liquid inherently adopts the characteristics of CFR, leading to an abrupt reduction in configurational entropy. This transition circumvents the emergence of the Kauzmann paradox and redirects the entropy-temperature curve away from the scenario predicted by Kauzmann. This also supports some similar arguments in the literature [67].

Finally, our previous study [22] was the first to identify two distinct types of supercooled liquids: GFR and CFR. In that work, we uncovered the difference in scaling behaviors and dynamic correlations between GFR and CFR. However, the specific differences in their PELs remain unclear. This study addresses this gap by demonstrating that the differences between the two supercooled liquids are not only dynamic in nature but also thermodynamic, particularly in terms of the accessible regions of their PELs.

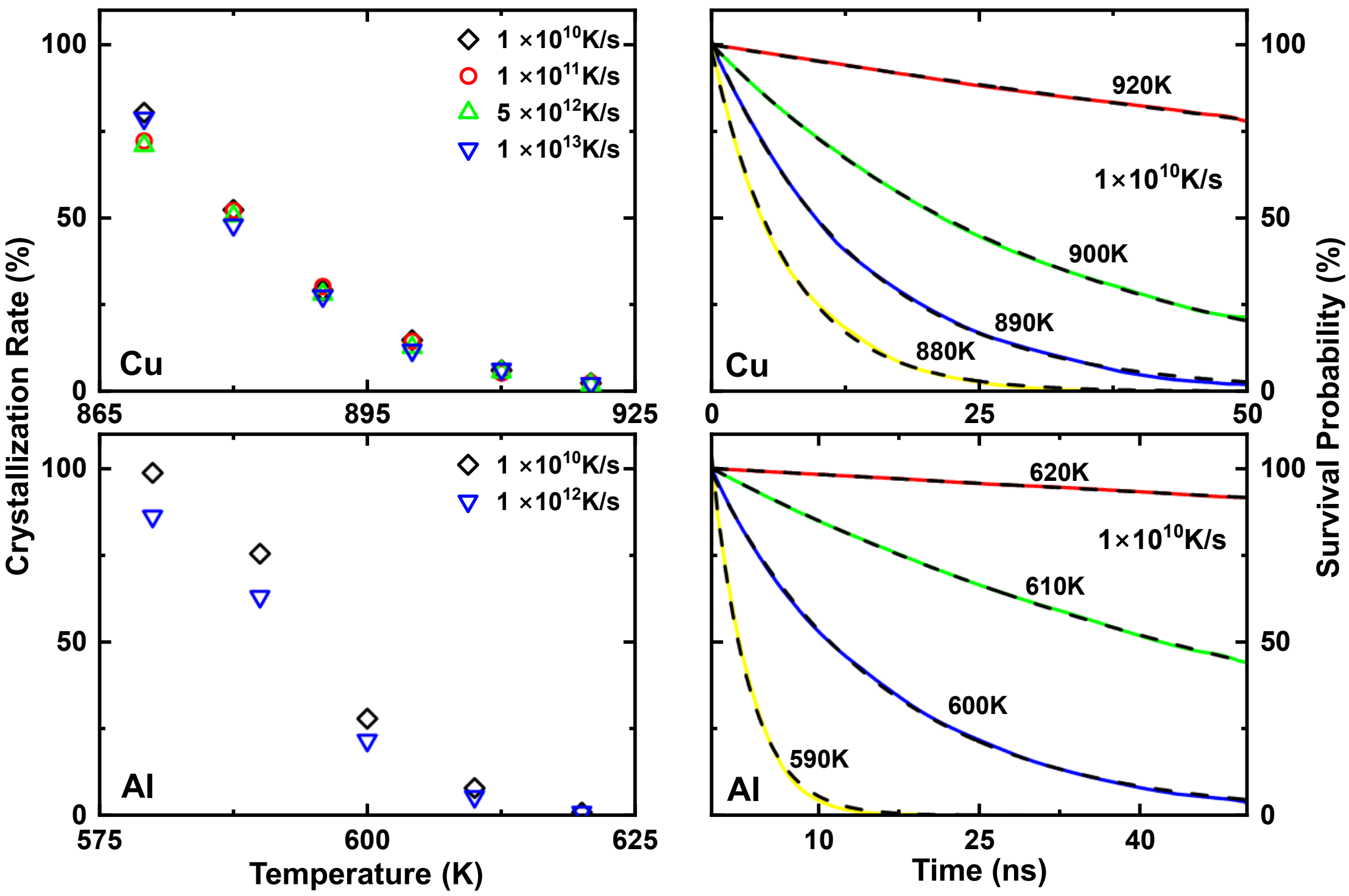

*Figure 4. Left panel: The crystallization rate ($P_c$) as a function of temperature for copper and aluminum with the crystallization observation window of 5ns. Right panel: The survival probability $Q(t)$ at a few selected temperatures for copper and aluminum. The dotted lines are the results fitted with exponential functions.*

### 3) Crystallization Dynamics

The left panel of Figure 4 illustrates $P_c$ for copper and aluminum, measured with $w_t = 5\text{ns}$. Notably, cooling rates exert only a modest influence on $P_c$ (see the left panel of Figure 4). Despite spanning three orders of magnitude in cooling rates, $P_c$ varies by less than 15% at the same temperatures. For all cooling rates studied, as expected, $P_c$ decreases with the increase of temperatures. A characteristic temperature $T_x$, approximately 890 K for copper and 615 K for aluminum, emerges as a pivotal threshold. Above $T_x$, $P_c$ for CFR and GFR is nearly indistinguishable. Below $T_x$, however, CFR exhibits consistently higher $P_c$ than GFR, indicating accelerated crystallization in CFR. Subsequent analysis reveals that $T_x$ coincides with a discontinuity in the temperature dependence of nucleation barriers (discussed below).

The right panel of Figure 4 presents the survival probability $Q(t)$ of supercooled liquids at selected temperatures. As the temperature decreases, $Q(t)$ decays more rapidly, reflecting shorter lifetimes $(\tau)$ of the metastable liquid. Fitting $Q(t))$ to an exponential decay function $(Q(t) = Ae^{-\frac{t}{\tau}})$ yields excellent agreement (dashed lines in Figure 4), validating the use of $\tau$ to quantify nucleation kinetics.

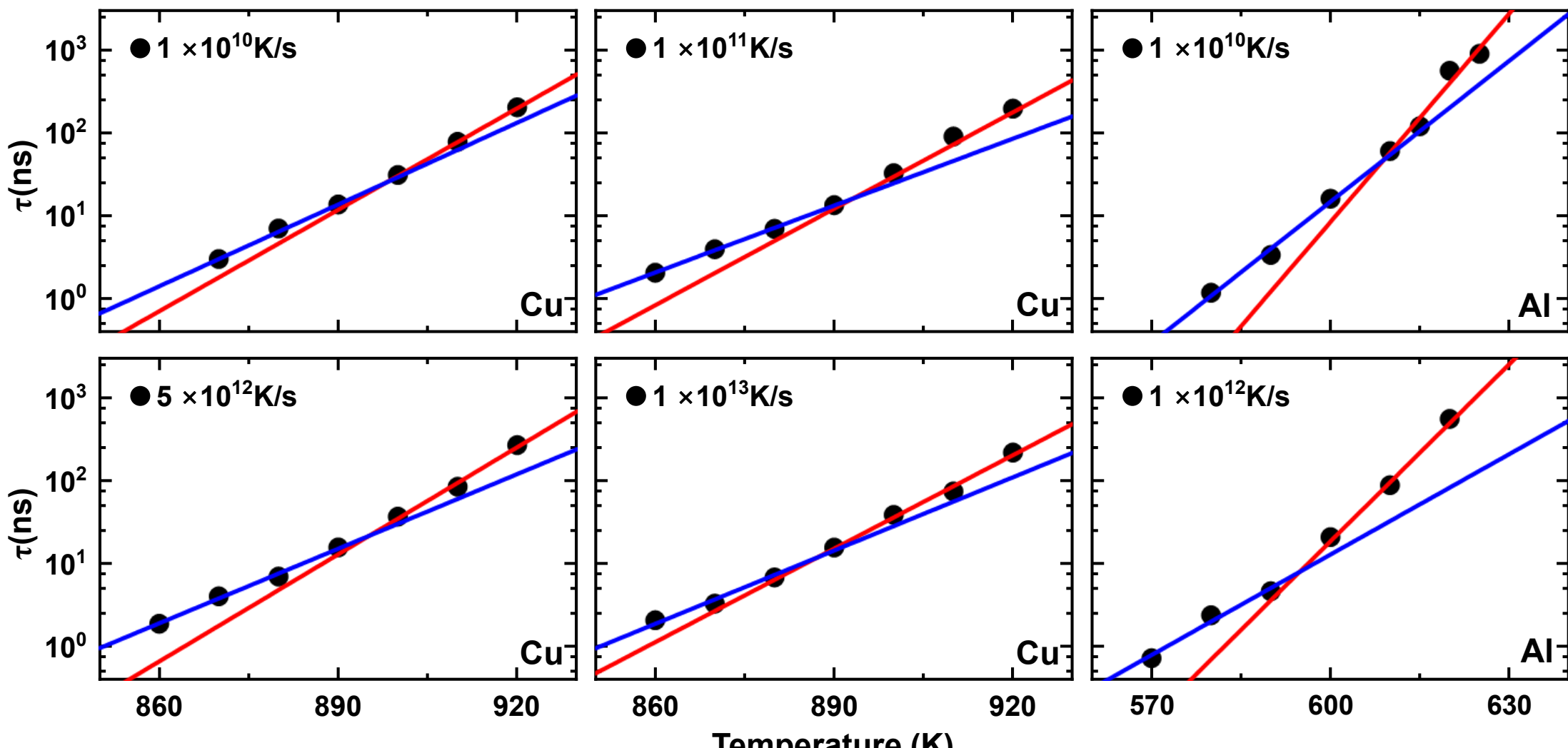

*Figure 5. Temperature dependence of the life time (τ). The vertical axis is plotted on a logarithmic scale. The logarithm of the life time is found to be proportional to temperature. A linear fitting with Eq. (1) is indicated by solid lines. The slopes at high and low temperatures are different, with the slope being precisely the parameter γ of Eq. (1).*

According to statistical mechanics, the life time $(\tau)$ of a supercooled liquid at temperature $T$ can be expressed as $\tau = \tau_0 exp\left(\frac{\Delta(T)}{k_B T}\right)$, where $\tau_0$ and $\Delta(T)$ are the characteristic time and nucleation barrier, respectively. Based on classical nucleation theory, $\Delta(T)$ is a function of temperature and decreases as temperature decreases [68, 69]. Generally, $\Delta(T)$ cannot be a linear function of temperature; otherwise, $\tau$ would not depend on temperature, which is clearly inconsistent with experimental observations. Considering only the leading term, we assume $\Delta(T) \approx k_B\sigma + \upsilon k_B T + \gamma k_B T^2 + O(T^3)$, where $\sigma$, $\upsilon$, and $\gamma$ are constants. Given that the temperatures of interest in this study are all in the high-temperature region, we can neglect the zero-order term ($\sigma$). MD results also support this approximation. If dropping $\sigma$, we obtain:

$$\ln\tau \approx \ln\tau_0 + \upsilon + \gamma T = \gamma_0 + \gamma T, \tag{1}$$

with $\gamma_0 = \ln\tau_0 + \upsilon$.

Figure 5 plots the logarithmic lifetime ($log_{10}\tau$) against temperature for both CFR and GFR. A linear relationship emerges, consistent with the theoretical expression of Eq. (1) (solid lines in Figure 5). The slope $\gamma$—linked to the temperature dependence of the nucleation barrier—differs markedly between high- and low-temperature regimes (Table I). Above $T_x$, $\gamma$ is remarkably larger, implying a steeper decrease in $\Delta(T)$ with temperature. This finding suggests that, supercooled liquids not only can be classified into two regimes (GFR and CFR), but each can also be divided into two sub-regimes with distinct crystallization behaviors, as shown in the left panel of Figure 1.

$T_x$ is only slightly changed as cooling rates varying, and very close to the freezing temperature $T_f$ (Table 1). Although current work is hard to establish a direct theoretical connection between $T_x$ and $T_f$, this bifurcation does underscore a fundamental shift in accessible regions of the PEL near $T_x$, likely tied to ergodicity-breaking transitions predicted by mode-coupling theory [70, 71].

Although $\gamma$ for GFR is larger than that for CFR, we cannot conclude that the crystallization barrier for GFR is lower. This is because $\gamma_0$ also differs between the two types of supercooled liquids, as shown in Table 1. Considering that both types of liquids belong to the same material class, $\tau_0$ should not differ by orders of magnitude. Thus, we can infer that $\upsilon$ differs considerably between the two types of supercooled liquids (note $\gamma_0 = \ln\tau_0 + \upsilon$). Considering the data shown in the left panel of Figure 4. we believe that the difference in the first-order term still ensures that CFR has a lower effective barrier in the temperature higher than $T_x$, despite the fact that the second-order term in $\Delta(T)$ changes more rapidly for GFR than for CFR.

*Table I. The obtained parameters using Eq. 1 fitting to the logarithmic lifetime (ln τ) against temperature.*

| | Cooling Rate(K/s) | High T $\gamma$ | Low T $\gamma$ | High T $\gamma_0$ | Low T $\gamma_0$ | $T_x$(K) | $T_f$(K) |
|---|---|---|---|---|---|---|---|
| Cu | $1 \times 10^{10}$ | 0.094 | 0.076 | -81.0 | -64.7 | 897 | 890 |
| | $1 \times 10^{11}$ | 0.089 | 0.062 | -76.7 | -52.4 | 883 | |
| | $5 \times 10^{12}$ | 0.099 | 0.069 | -85.5 | -58.6 | 895 | |
| | $1 \times 10^{13}$ | 0.086 | 0.068 | -74.2 | -57.6 | 887 | |
| Al | $1 \times 10^{10}$ | 0.193 | 0.130 | -113.5 | -74.5 | 609 | 600 |
| | $1 \times 10^{12}$ | 0.164 | 0.093 | -95.3 | -53.2 | 595 | |

## Summary

From both thermodynamic and dynamic perspectives, we systematically investigated the differences between supercooled liquids formed above CCR and below CCR. Our findings reveal that the distribution of the average potential energies in CFR is notably steep, whereas that of GFR is significantly broader. Consequently, we suggest that CFR possesses a lower configurational entropy. By comparing the crystallization rates and relaxation times of the two types of supercooled liquids, we have identified a characteristic temperature $T_x$ for both liquids, at which the relationship between the nucleation barrier and temperature undergoes a discontinuous change. This implies that within a specific type of supercooled liquid, the system can be delineated into two distinct regions, high temperature and low temperature regions, based on its crystallization barrier. Furthermore, $T_x$ is close to the freezing temperature $T_f$.

**Acknowledgements:** This work was supported by the National Key Research and Development Program of China (Grant No. 2022YFA1404603), and by the National Natural Science Foundation of China (Grant No. 12274127 and 12188101).

## References

[1] E. Asayama, H. Takebe, K. Morinaga, Critical Cooling Rates for the Formation of Glass for Silicate Melts, Isij Int, 33 (1993) 233-238.

[2] S. Guo, Y. Liu, Estimation of critical cooling rates for formation of amorphous alloys from critical sizes, J Non-Cryst Solids, 358 (2012) 2753-2758.

[3] N. Liu, T. Ma, C. Liao, G. Liu, R.M.O. Mota, J. Liu, S. Sohn, S. Kube, S. Zhao, J.P. Singer, J. Schroers, Combinatorial measurement of critical cooling rates in aluminum-base metallic glass forming alloys, Sci Rep, 11 (2021) 3903.

[4] M.P. Hazarika, P. Bordoloi, A. Tripathi, S.N. Chakraborty, Understanding crystallization and amorphization in liquid Ti cooled at different rates: A molecular dynamics simulation study, J Chem Phys, 161 (2024) 234503.

[5] Q. Zhang, J. Wang, S. Tang, Y. Wang, J. Li, W. Zhou, Z. Wang, Molecular dynamics investigation of the local structure in iron melts and its role in crystal nucleation during rapid solidification, Phys Chem Chem Phys, 21 (2019) 4122-4135.

[6] M. Shimono, H. Onodera, Molecular dynamics study on formation and crystallization of Ti–Al amorphous alloys, Mater Sci Eng A, 304-306 (2001) 515-519.

[7] Z.-A. Tian, R.-S. Liu, H.-R. Liu, C.-X. Zheng, Z.-Y. Hou, P. Peng, Molecular dynamics simulation for cooling rate dependence of solidification microstructures of silver, J Non-Cryst Solids, 354 (2008) 3705-3712.

[8] A. Takeuchi, A. Inoue, Quantitative evaluation of critical cooling rate for metallic glasses, Mater Sci Eng A, 304-306 (2001) 446-451.

[9] C.S. Liu, J. Xia, Z.G. Zhu, D.Y. Sun, The cooling rate dependence of crystallization for liquid copper: A molecular dynamics study, J Chem Phys, 114 (2001) 7506-7512.

[10] P.G. Debenedetti, F.H. Stillinger, Supercooled liquids and the glass transition, Nature, 410 (2001) 259-267.

[11] D. Turnbull, Under what conditions can a glass be formed?, Contemp Phys, 10 (1969) 473-488.

[12] J.G. Wang, On the formation of metallic glass, J Non-Cryst Solids, 649 (2025) 123329.

[13] D.V. Louzguine-Luzgin, M. Miyama, K. Nishio, A.A. Tsarkov, A.L. Greer,

Vitrification and nanocrystallization of pure liquid Ni studied using molecular-dynamics simulation, J Chem Phys, 151 (2019) 124502.

[14] S. Assouli, H. Jabraoui, T. El Hafi, O. Bajjou, A. Kotri, M. Mazroui, Y. Lachtioui, Exploring the impact of cooling rates and pressure on fragility and structural transformations in iron monatomic metallic glasses: Insights from molecular dynamics simulations, J Non-Cryst Solids, 621 (2023) 122623.

[15] T. El hafi, O. Bajjou, H. Jabraoui, J. Louafi, M. Mazroui, Y. Lachtioui, Effects of cooling rate on the glass formation process and the microstructural evolution of Silver mono-component metallic glass, Chem Phys, 569 (2023) 111873.

[16] N. Clavaguera, Non-equilibrium crystallization, critical cooling rates and transformation diagrams, J Non-Cryst Solids, 162 (1993) 40-50.

[17] L. Qi, H.F. Zhang, Z.Q. Hu, Molecular dynamic simulation of glass formation in binary liquid metal: Cu-Ag using EAM, Intermetallics, 12 (2004) 1191-1195.

[18] L. Zhong, J. Wang, H. Sheng, Z. Zhang, S.X. Mao, Formation of monatomic metallic glasses through ultrafast liquid quenching, Nature, 512 (2014) 177-180.

[19] J.C. Mauro, P.K. Gupta, R.J. Loucks, Continuously broken ergodicity, J Chem Phys, 126 (2007) 184511.

[20] J.C. Mauro, M.M. Smedskjaer, Statistical mechanics of glass, J Non-Cryst Solids, 396 (2014) 41-53.

[21] J.C. Mauro, R.J. Loucks, S. Sen, Heat capacity, enthalpy fluctuations, and configurational entropy in broken ergodic systems, J Chem Phys, 133 (2010) 164503.

[22] B. Zhang, D.M. Zhang, D.Y. Sun, X.G. Gong, Scaling in Kinetics of Supercooled Liquids, under review; arXiv:2308.09572v3.

[23] A. Heuer, Properties of a Glass-Forming System as Derived from Its Potential Energy Landscape, Phys Rev Lett, 78 (1997) 4051-4054.

[24] F.H. Stillinger, A topographic view of supercooled liquids and glass formation, Science, 267 (1995) 1935-1939.

[25] F. Bamer, F. Ebrahem, B. Markert, B. Stamm, Molecular Mechanics of Disordered Solids, Arch Comput Method E, 30 (2023) 2105-2180.

[26] F.H. Stillinger, P.G. Debenedetti, Phase transitions, Kauzmann curves, and inverse

melting, Biophys Chem, 105 (2003) 211-220.

[27] F.H. Stillinger, Supercooled liquids, glass transitions, and the Kauzmann paradox, The Journal of Chemical Physics, 88 (1988) 7818-7825.

[28] W. Kauzmann, The Nature of the Glassy State and the Behavior of Liquids at Low Temperatures, Chem Rev, 43 (2002) 219-256.

[29] Y. Ashkenazy, R.S. Averback, Kinetic stages in the crystallization of deeply undercooled body-centered-cubic and face-centered-cubic metals, Acta Mater, 58 (2010) 524-530.

[30] K.F. Kelton, Crystal Nucleation in Liquids and Glasses, Solid State Phys, 45 (1991) 75-177.

[31] D.W. Oxtoby, Homogeneous Nucleation - Theory and Experiment, J Phys-Condens Mat, 4 (1992) 7627-7650.

[32] W. Klein, F. Leyvraz, Crystalline Nucleation in Deeply Quenched Liquids, Phys Rev Lett, 57 (1986) 2845-2848.

[33] M.J. Mandell, J.P. Mctague, A. Rahman, Crystal Nucleation in a 3-Dimensional Lennard-Jones System - Molecular-Dynamics Study, J Chem Phys, 64 (1976) 3699-3702.

[34] E.D. Zanotto, J.C. Mauro, The glassy state of matter: Its definition and ultimate fate, J Non-Cryst Solids, 471 (2017) 490-495.

[35] J.S. Vanduijneveldt, D. Frenkel, Computer-Simulation Study of Free-Energy Barriers in Crystal Nucleation, J Chem Phys, 96 (1992) 4655-4668.

[36] C. Desgranges, J. Delhommelle, Unusual Crystallization Behavior Close to the Glass Transition, Phys Rev Lett, 120 (2018) 115701.

[37] J. Bokeloh, G. Wilde, R.E. Rozas, R. Benjamin, J. Horbach, Nucleation barriers for the liquid-to-crystal transition in simple metals: Experiment vs. simulation, Eur Phys J-Spec Top, 223 (2014) 511-526.

[38] M. Goldstein, Viscous liquids and the glass transition: a potential energy barrier picture, J Chem Phys, 51 (1969) 3728-3739.

[39] P.R. tenWolde, M.J. RuizMontero, D. Frenkel, Numerical calculation of the rate of crystal nucleation in a Lennard-Jones system at moderate undercooling, J Chem Phys,

104 (1996) 9932-9947.

[40] M.I. Mendelev, M.J. Kramer, C.A. Becker, M. Asta, Analysis of semi-empirical interatomic potentials appropriate for simulation of crystalline and liquid Al and Cu, Philos Mag, 88 (2008) 1723-1750.

[41] F. Ercolessi, J.B. Adams, Interatomic potentials from first-principles calculations: the force-matching method, Europhys Lett, 26 (1994) 583-588.

[42] A.P. Thompson, H.M. Aktulga, R. Berger, D.S. Bolintineanu, W.M. Brown, P.S. Crozier, P.J. in 't Veld, A. Kohlmeyer, S.G. Moore, T.D. Nguyen, R. Shan, M.J. Stevens, J. Tranchida, C. Trott, S.J. Plimpton, LAMMPS - a flexible simulation tool for particle-based materials modeling at the atomic, meso, and continuum scales, Comput Phys Commun, 271 (2022) 108171.

[43] L.R. Medrano, C.V. Landauro, J. Rojas-Tapia, Implementation of an alternative method to determine the critical cooling rate: Application in silver and copper nanoparticles, Chem Phys Lett, 612 (2014) 273-279.

[44] J. Liu, J.Z. Zhao, Z.Q. Hu, The development of microstructure in a rapidly solidified Cu, Mater Sci Eng A, 452-453 (2007) 103-109.

[45] S.A. Rogachev, A.S. Rogachev, M.I. Alymov, Estimating the Critical Glass Transition Rate of Pure Metals Using Molecular Dynamic Modeling, Doklady Physics, 64 (2019) 214-217.

[46] D.M. Zhang, D.Y. Sun, X.G. Gong, Discovery of a paired Gaussian and long-tailed distribution of potential energies in nanoglasses, Phys Rev B, 105 (2022) 035403.

[47] F.H. Stillinger, T.A. Weber, Hidden Structure in Liquids, Phys Rev A, 25 (1982) 978-989.

[48] S. Sastry, P.G. Debenedetti, F.H. Stillinger, Signatures of distinct dynamical regimes in the energy landscape of a glass-forming liquid, Nature, 393 (1998) 554-557.

[49] Y. Nishikawa, M. Ozawa, A. Ikeda, P. Chaudhuri, L. Berthier, Relaxation Dynamics in the Energy Landscape of Glass-Forming Liquids, Phys Rev X, 12 (2022) 021001.

[50] F. Sciortino, Potential energy landscape description of supercooled liquids and glasses, J Stat Mech-Theory E, DOI (2005) P05015.

[51] A. Heuer, Exploring the potential energy landscape of glass-forming systems: from inherent structures via metabasins to macroscopic transport, J Phys Condens Matter, 20 (2008) 373101.
[52] M. Baity-Jesi, G. Biroli, D.R. Reichman, Revisiting the concept of activation in supercooled liquids, Eur Phys J E, 44 (2021) 77.
[53] F.H. Stillinger, T.A. Weber, Packing Structures and Transitions in Liquids and Solids, Science, 225 (1984) 983-989.
[54] K. Gonzalez-Lopez, E. Lerner, An energy-landscape-based crossover temperature in glass-forming liquids, J Chem Phys, 153 (2020) 241101.
[55] T.S. Grigera, A. Cavagna, I. Giardina, G. Parisi, Geometric approach to the dynamic glass transition, Phys Rev Lett, 88 (2002) 055502.
[56] B. Doliwa, A. Heuer, Hopping in a supercooled Lennard-Jones liquid: Metabasins, waiting time distribution, and diffusion, Phys Rev E, 67 (2003) 030501.
[57] J. Tan, S.R. Zhao, W.F. Wang, G. Davies, X.X. Mo, The effect of cooling rate on the structure of sodium silicate glass, Mat Sci Eng B-Solid, 106 (2004) 295-299.
[58] W. Sun, V. Dierolf, H. Jain, Molecular dynamics simulation of the effect of cooling rate on the structure and properties of lithium disilicate glass, J Non-Cryst Solids, 569 (2021) 120991.
[59] W.C. Huang, H. Jain, E.I. Kamitsos, A.P. Patsis, Anomalous expansion of sodium triborate melt and its effect on glass properties, J Non-Cryst Solids, 162 (1993) 107-117.
[60] R. Bruning, M. Sutton, Volume of B2O3 at the glass transition, Phys Rev B Condens Matter, 49 (1994) 3124-3130.
[61] Y.Z. Yue, R. von der Ohe, S.L. Jensen, Fictive temperature, cooling rate, and viscosity of glasses, J Chem Phys, 120 (2004) 8053-8059.
[62] S. Sarkar, B. Bagchi, Inherent structures of phase-separating binary mixtures: Nucleation, spinodal decomposition, and pattern formation, Phys Rev E, 83 (2011) 031506.
[63] C. Rehwald, N. Gnan, A. Heuer, T. Schroder, J.C. Dyre, G. Diezemann, Aging effects manifested in the potential-energy landscape of a model glass former, Phys Rev

E, 82 (2010) 021503.

[64] I. Saika-Voivod, F. Sciortino, Distributions of inherent structure energies during aging, Phys Rev E, 70 (2004) 041202.

[65] J. Chowdhary, T. Keyes, Thermodynamics and dynamics for a model potential energy landscape, J Phys Chem B, 108 (2004) 19786-19798.

[66] F. Sciortino, W. Kob, P. Tartaglia, Thermodynamics of supercooled liquids in the inherent-structure formalism: a case study, J Phys-Condens Mat, 12 (2000) 6525-6534.

[67] F.H. Stillinger, P.G. Debenedetti, T.M. Truskett, The Kauzmann paradox revisited, J Phys Chem B, 105 (2001) 11809-11816.

[68] A. Ziabicki, L. Jarecki, Potential energy barriers in the kinetic theory of nucleation, J Chem Phys, 80 (1984) 5751-5753.

[69] S. Ayuba, D. Suh, K. Nomura, T. Ebisuzaki, K. Yasuoka, Kinetic analysis of homogeneous droplet nucleation using large-scale molecular dynamics simulations, J Chem Phys, 149 (2018) 044504.

[70] D.R. Reichman, P. Charbonneau, Mode-coupling theory, Journal of Statistical Mechanics: Theory and Experiment, 2005 (2005) P05013.

[71] W. Götze, Recent tests of the mode-coupling theory for glassy dynamics, Journal of Physics: Condensed Matter, 11 (1999) A1-A45.